\documentclass[11pt,twoside]{article}
\usepackage{asp2010}

\def\teff{T_{\rm eff}}

\def\kms{km/s}

\def\halpha{H$\alpha$}

\def\synthmag{{\sc Synthmag}}
\def\synmast{{\sc Synmast}}

\def\vsini{$\upsilon\sin i$}

\def\hunda{Hund's case~(a)}
\def\hundb{Hund's case~(b)}

\def\b{|\mathrm{\mathbf{B}}|}
\def\bv{\mathrm{\mathbf{B}}}
\def\btimesf{(|\mathrm{\mathbf{B}}|f)}

\resetcounters

\begin{document}

\title{Magnetic fields in M-dwarfs: quantitative results from detailed
  spectral synthesis in FeH lines} \author{Shulyak, D.$^1$, Reiners,
  A.$^1$, Wende, S. $^1$, Kochukhov, O.$^2$, Piskunov, N.$^2$,
  Seifahrt, A.$^3$ \affil{$^1$Universit\"at G\"ottingen, Institut
    f\"ur Astrophysik, Friedrich-Hund-Platz 1, D-37077 G\"ottingen,
    Germany} \affil{$^2$Department of Physics and Astronomy, Uppsala
    University, Box 516, 751 20, Uppsala, Sweden}
  \affil{$^3$Department of Physics, University of California, One
    Shields Avenue, Davis, CA 95616, USA} }

\begin{abstract}
  Strong surface magnetic fields are ubiquitously found in M-dwarfs
  with mean intensities on the order of few thousand Gauss~--~three
  orders of magnitude higher than the mean surface magnetic field of
  the Sun.  These fields and their interaction with photospheric
  convection are the main source of stellar activity, which is of big
  interest to study links between parent stars and their planets.
  Moreover, the understanding of stellar magnetism, as well as the
  role of different dynamo-actions in particular, is impossible
  without explaining magnetic fields in M-dwarfs. Measuring magnetic
  field intensities and geometries in such cool objects, however, is
  strongly limited to our ability to simulate the Zeeman effect in
  molecular lines.  In this work, we present quantitative results of
  modelling and analysis of the magnetic fields in selected M-dwarfs
  in FeH Wing-Ford lines and strong atomic lines. Some particular FeH
  lines are found to be the excellent probes of the magnetic field.
\end{abstract}

\section{Introduction}
Magnetic fields in non-degenerate stars are found all across the
Hertzsprung-Russell diagram, from hot high-luminous stars down to cool
and ultra-cool dwarfs \citep[see, for example, the review by][and
references therein]{donati-landstreet2009}.
The characterization of the magnetic fields in cool low-mass M-stars
is of high interest because a) these stars contribute up to 70\% by number of the total
stellar population of our galaxy; b)  these stars
often show dynamo-generated activity in their atmospheres similar to that of the Sun 
(i.e. with flares seen as strong emission in X-ray, \halpha, Ca~II~H~\&~K lines, see \citet{berdyugina2005}), 
however, objects later than M$3.5$ ($\teff \approx 3400$K, M$<0.35$M$_{\odot}$) are believed to become fully convective, therefore different dynamo mechanisms
need to be involved to explain their fields; c) these stars are promising targets to search for Earth-size planets inside their
habitable zones, and thus knowledge of the magnetic activity of a parent star is potentially important for the climate modelling
of these distant earth's. In this paper we will present first quantitative and independent estimates of the surface magnetic fields
in selected M-dwarfs based on synthetic spectra modeling in FeH lines of the Wing-Ford $F^4\,\Delta-X^4\,\Delta$ transitions.

\section{What is known about magnetic fields of M-dwarf up to now?}

Measurements of the magnetic fields in cool stars relies on
polarimetric observations and analysis of Zeeman broadening of spectral lines. 
Utilizing  the  Zeeman Doppler Imaging (ZDI) and Least Square Deconvolution (LSD)
techniques as applied to Stokes $V$ spectra, \citet{donati2008} and \citet{morin2008}
carried out a systematic characterization of the magnetic field topologies in a sample
of M-dwarfs. Authors found a clear transition in the magnetic field topologies between partly and fully convective stars, 
with later tend to host strong, poloidal, mostly axisymmetric fields contrary to much weaker, non-axisymmetric fields with strong toroidal component
for the former. In contrast, most recent investigation revealed a presence of M-dwarfs
with strong toroidal non-axisymmetric fields among fully convective objects, and thus no clear transition between these two
groups of M-stars seem to exist any more \citep{morin2010}. These results must be taken with certain caution, since using LSD technique
and limiting only to Stokes $V$ may result in missing an important information about the true magnetic field geometry and
total magnetic flux, as discussed in the talk of A.~Reiners (this meeting) and recent study by \cite{kochukhov2010}.

The Zeeman effect is sensitive to the surface averaged magnetic field
modulus and thus naturally gives information about the true magnetic field flux.
Strong fields up to $\approx4$~kG were then reported for some M-dwarfs based
on the relative analysis of magnetically sensitive Fe~I line at $\lambda=8467$\AA\,\citep{jk-valenti1996,jk-valenti2000}. 
For dwarfs cooler than mid~-~M, 
molecular lines of FeH Wing-Ford $F^4\,\Delta-X^4\,\Delta$ transitions around
$0.99~\mu$m are usually utilized \citep{valenti2001,r-and-b-2006}.
Some of these lines do show strong magnetic sensitivity, as seen, for instance, in the
sunspot spectra \citep{wallace1998}.

\section{Land\'e g-factors of FeH: roots of the problem}

Unfortunately, most lines of FeH are formed in the intermediate Hund's
case, which theoretical description is based on certain approximations.
The main problem is connected with the Born-Oppenheimer approximation,
which assumes a clear separation between the electronic and the nuclear motion in terms of energies.
This approximation fails for FeH, and no satisfactory description of Land\'e g-factors exist so far.
Among recent improvements in understanding the Zeeman splitting in diatomic molecules one should mention papers by \citet{berdyugina2002} 
and \citet{ramos2006}. However, theoretical g-factors still cannot describe correctly the broadening of great majority of FeH lines.

An alternative solution was then suggested by
\citet{r-and-b-2006,r-and-b-2007}, who estimated the $\btimesf$ values ($f$--filling factor)
 in a number of M-dwarfs by simple linear interpolation between
the spectral features of two reference stars, one with zero field and another one
with the magnetic field known from \citet{jk-valenti1996}, but the error bars of such an analysis stays high ($\approx1$~kG).

\begin{figure}
{\includegraphics[height=0.5\hsize,width=0.163\vsize,angle=-90]{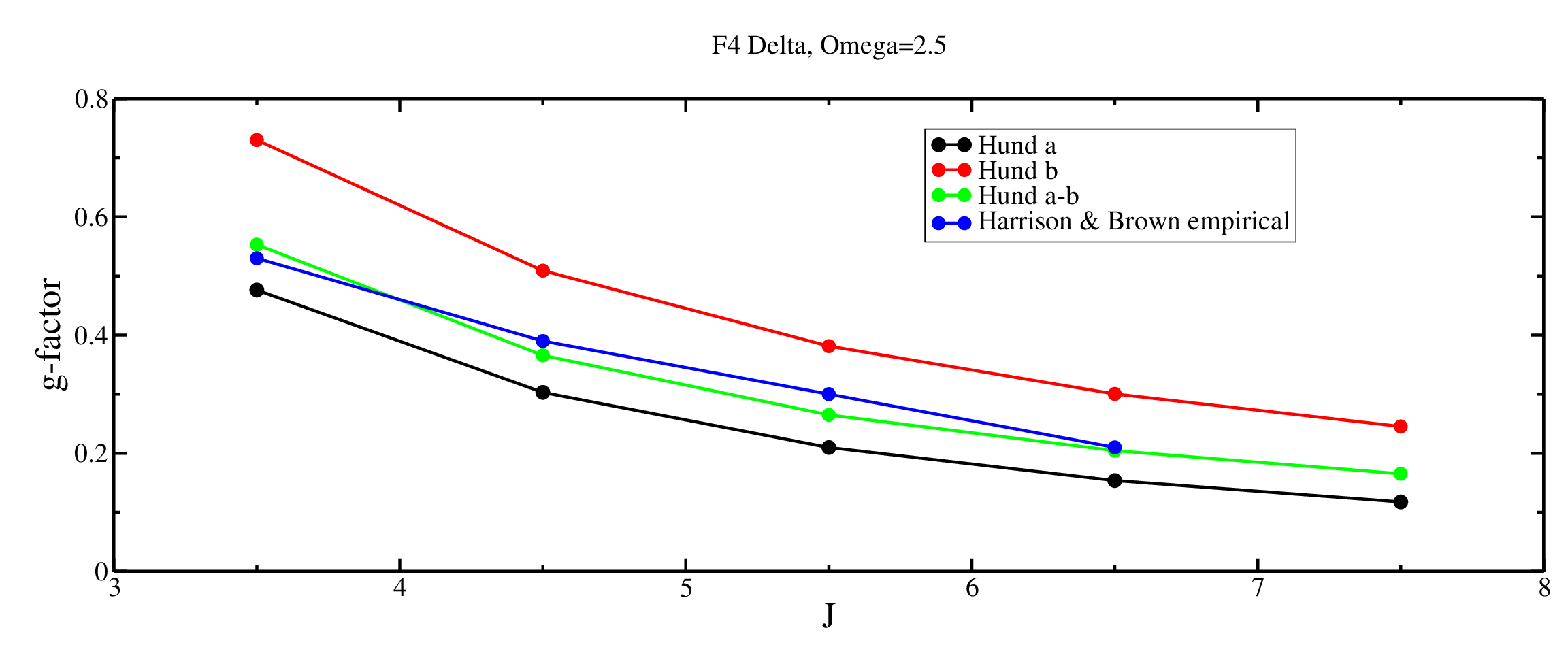}}
{\includegraphics[height=0.5\hsize,angle=-90]{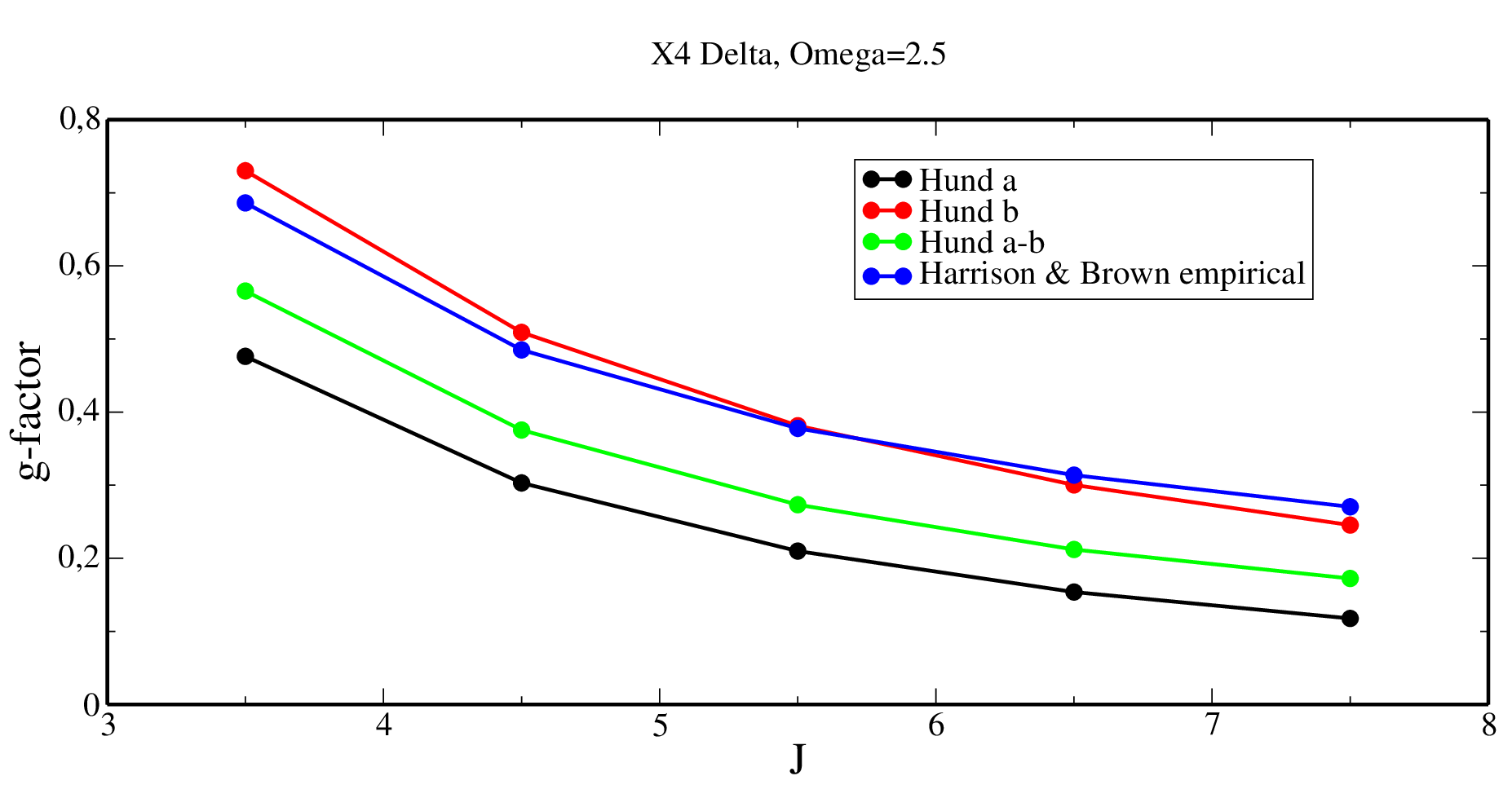}}
\caption{Comparison between observed \citep{harrison-brown2008} and theoretical
g-factors of lower ($X^4~\Delta$) and upper ($F^4~\Delta$) levels of FeH with rotation quantum number $\Omega=2.5$ 
as a function of magnetic quantum number $J$. Black line~--~ pure Hund's case (a),
red line~--~pure Hund's case (b), green line~--~intermediate between pure Hund's (a) and (b) cases.}
\label{fig:h-and-b}
\end{figure}

First semi-empirical fit to FeH lines in a sunspot spectra was presented
in \citet{afram2008}.
The authors succeeded to obtained a very good agreement with
observations for selected FeH lines and presented best-fitted
polynomial g-factors of upper and lower levels of corresponding
transitions. A little earlier, \citet{harrison-brown2008} presented
measured g-factors for a number of FeH lines, but were
limited to low magnetic $J$-numbers only.
As an example, Fig.~\ref{fig:h-and-b} illustrates a comparison between
experimental and theoretical g-factors for the lower ($X^4\,\Delta$) and
upper ($F^4\,\Delta$) states of FeH transitions with rotational quantum number $\Omega=2.5$.
It is seen that the lower state tends to be close to the pure Hund's case (b) and the upper
state is in intermediate case, while theoretical predictions always give solutions between
pure (a) and (b) cases, thus failing to reproduce g-factors for the lower state.

Below we attempt to provide an independent
measurement from a number of magnetically very sensitive molecular FeH
lines, which are modelled based on the formalism described in \citet{berdyugina2002}.

\section{Methods}

The line list of FeH transitions and molecular constants were taken from \citet{dulick2003}\footnote{http://bernath.uwaterloo.ca/FeH/}
and corrected as described in \citet{wende2010}.
The VALD (Vienna Atomic Line Database) was used as a source of atomic transitions \citep{vald1,vald2}.
Magnetic spectra synthesis is performed using the \synmast\ code \citep{synthmag2007}. The code
represents an improved version of the \synthmag\ code described by
\citet{synthmag}. The model atmospheres are from the recent MARCS grid\footnote{http://marcs.astro.uu.se} \citep{marcs}.
To compute g-factors in different Hund's cases, we implemented numerical libraries from the MZL (Molecular Zeeman Library)
package originally written by B.~Leroy \citep{mzl}, and adopted
for the particular case of FeH.

\section{Results}

A sunspot spectrum is probably the only trustworthy data source to test the predicted Zeeman patterns of FeH lines because 
a) the temperature inside a spot is still hot enough to see strong unblended atomic and FeH lines and 
b) very high-resolution and S/N observations are available. We thus made use of an umbra spectrum
from \citet{wallace1998}. They also derived the magnetic field intensity $\b=2.7$~kG.
Our fit to the atomic lines (mostly Fe, Ti, and Cr) in the range $9800-10\,800$\,\AA\ also 
suggests a field intensity $\b=2.7$~kG (model atmosphere with $\teff=4000$~K). 
%As an example, Fig.~\ref{fig:spot-atoms} illustrates theoretical fits for some Fe~I lines.
%
%\begin{figure}
%\centerline{\includegraphics[width=0.7\hsize]{figures/15229fg1a.ps}}
%%\centerline{\includegraphics[width=0.7\hsize]{figures/15229fg1b.ps}}
%\caption{Observed and predicted spectra of a sunspot. Theoretical computations are shown for two
%magnetic field geometries: purely radial ($\br,0,0$) and with horizontal contribution ($\br,\bm,0$)
%(see legends on individual plots). Wavelengths are in vacuum.}
%\label{fig:spot-atoms}
%\end{figure}

\begin{figure*}[ht]
\includegraphics[width=\hsize]{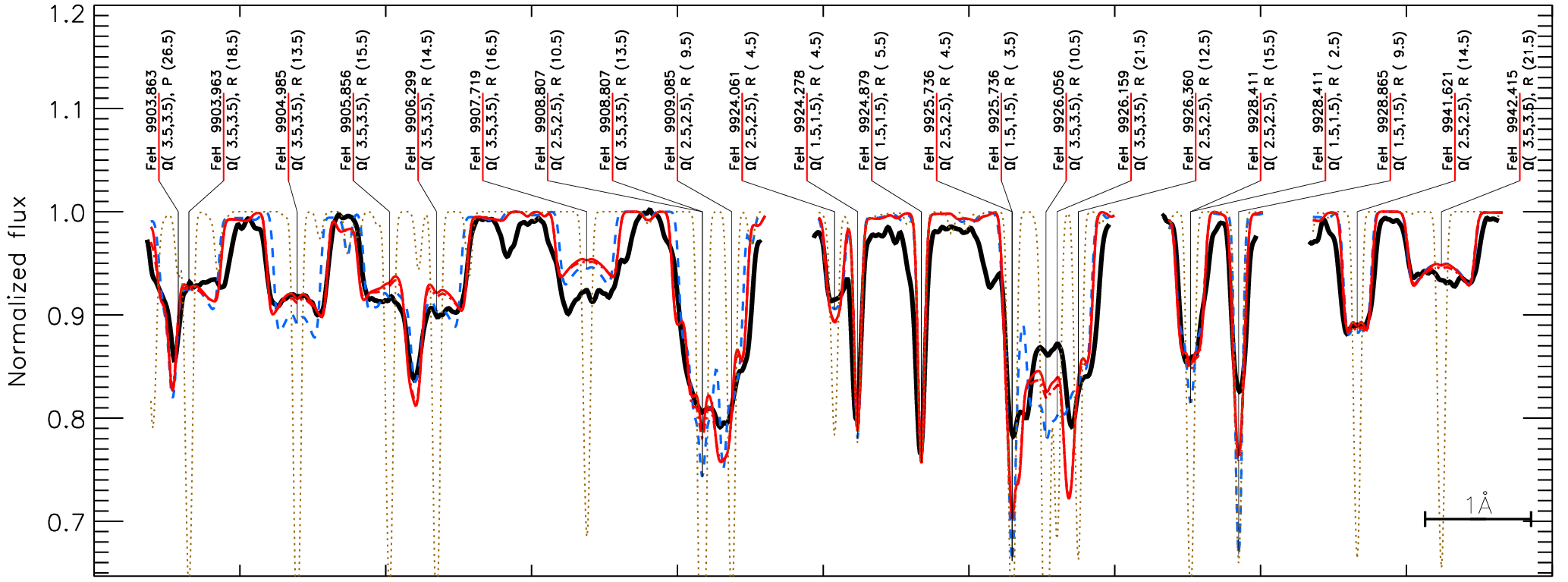}
\includegraphics[width=\hsize]{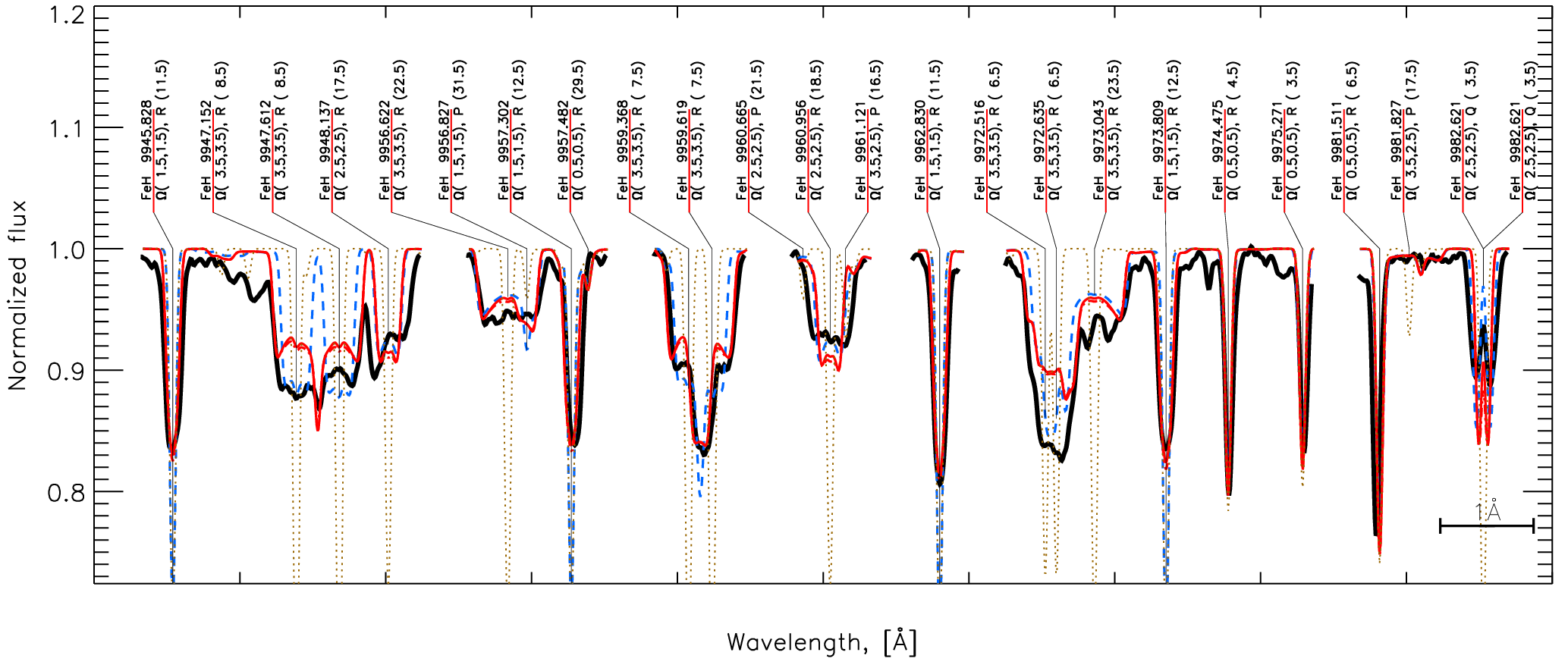}
\caption{Comparison between observed and theoretical sunspot spectra in selected regions of FeH transitions. 
Blue dashed line~--~best-fitted g-factor from \citet{afram2008}, red solid line~--~calculations with MZL library; 
both with a purely radial field of $\bv=(2.7,0,0)$~kG. Red dash-dotted line~--~ synthetic spectra accounted for the horizontal field 
component $\bv=(2.5,1,0)$~kG (hardly seen in the figure, coincides with red solid line),
brown dotted line~--~zero-field spectrum.
Labels over lines indicate their central wavelengths, omegas of lower and upper states (in brackets), branch, and the $J$-number
of the lower state (in brackets). Wavelengths are in vacuum.}
\label{fig:spot-feh}
\end{figure*}

Using the MZL library and sunspot spectra we find that the intermediate Hund's case (with its present treatment in MZL) 
is a good approximation if  ($l$~--~lower, $u$~--~upper states)
\begin{enumerate}
\item
$\Omega_l=0.5$
\item
$\Omega_{l \; \mathrm{or} \; u}\leq2.5$ and $3Y>J(J+1)$ for the P and Q branches
\item
$\Omega_{l \; \mathrm{and} \; u}=2.5$ and $5Y>J(J+1)$ for the R branch.
\end{enumerate}
For the rest of transitions the assumption of \hunda\ for the upper level 
and \hundb\ for lower level provide reasonable results.

Figure~\ref{fig:spot-feh} illustrates a comparison between best-fitted g-factor from \citet{afram2008} and our calculation for
some selected lines in the $9900-10\,000$~\AA\ region. Note that apart from the figures shown in \citet{afram2008}, we did not make an attempt
of correcting the theoretical spectra, i.e. no filling factors were applied. The model atmosphere and field intensity are the same as
determined previously from the metallic line spectra. 
Still, the discrepancy is found for low omega R-branch lines like FeH $9945$~\AA, $9962$~\AA, etc., which indeed show splitting closer to pure Hund's cases.
Even though we succeeded well enough to fit the width of the observed lines with the same field of $\b=2.7$~kG derived previously from atomic lines,
this cannot be judged to be more physical though until new improvements in the theoretical description of the intermediate case
will become available.

\begin{figure*}
\includegraphics[width=\hsize]{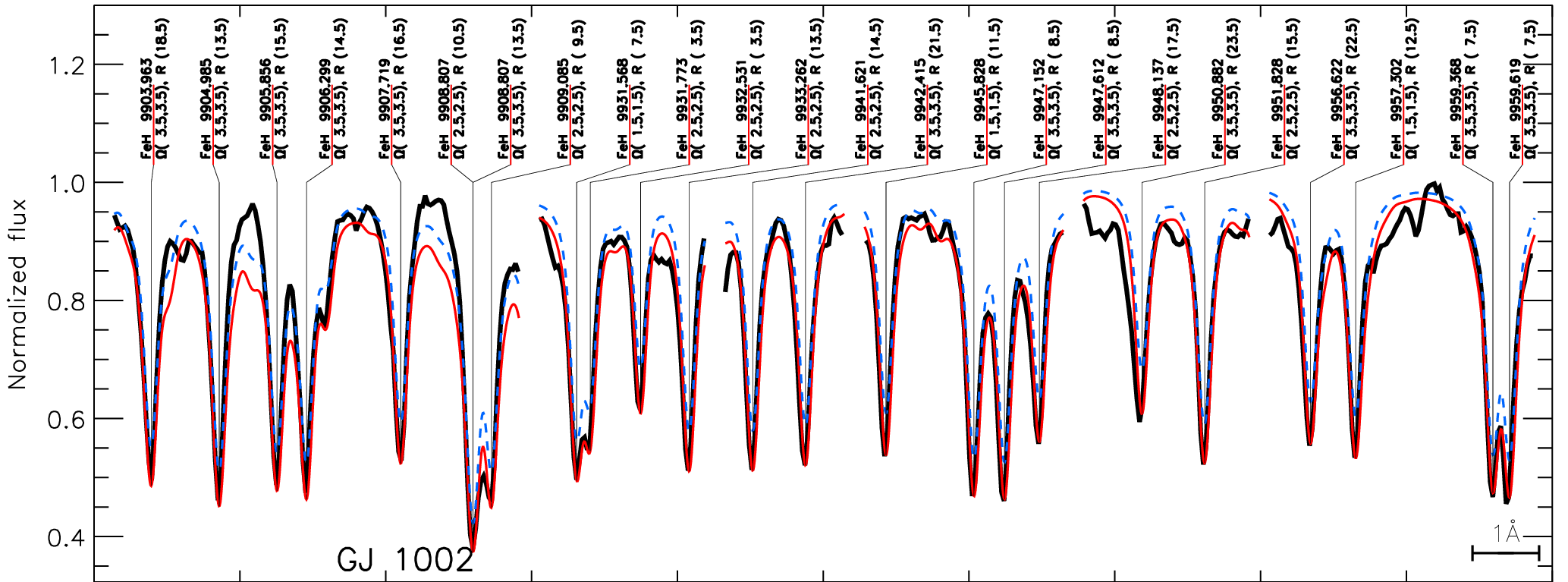}
\includegraphics[width=\hsize]{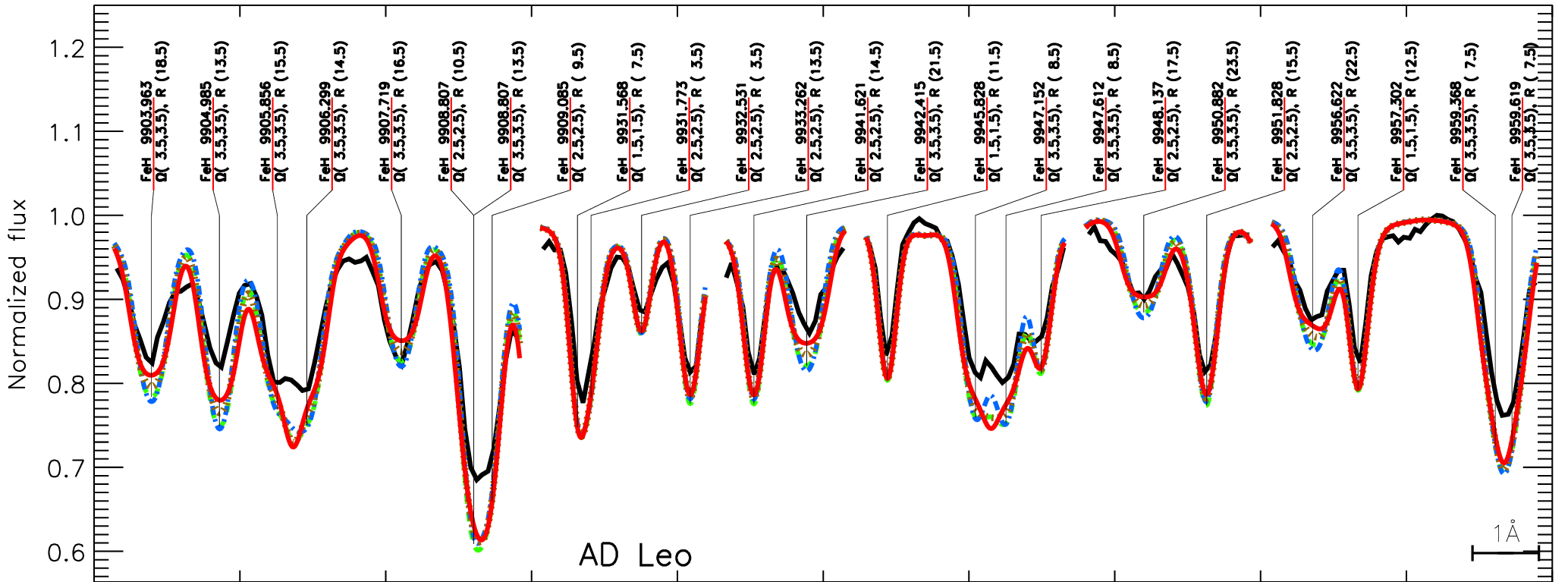}
\includegraphics[width=\hsize]{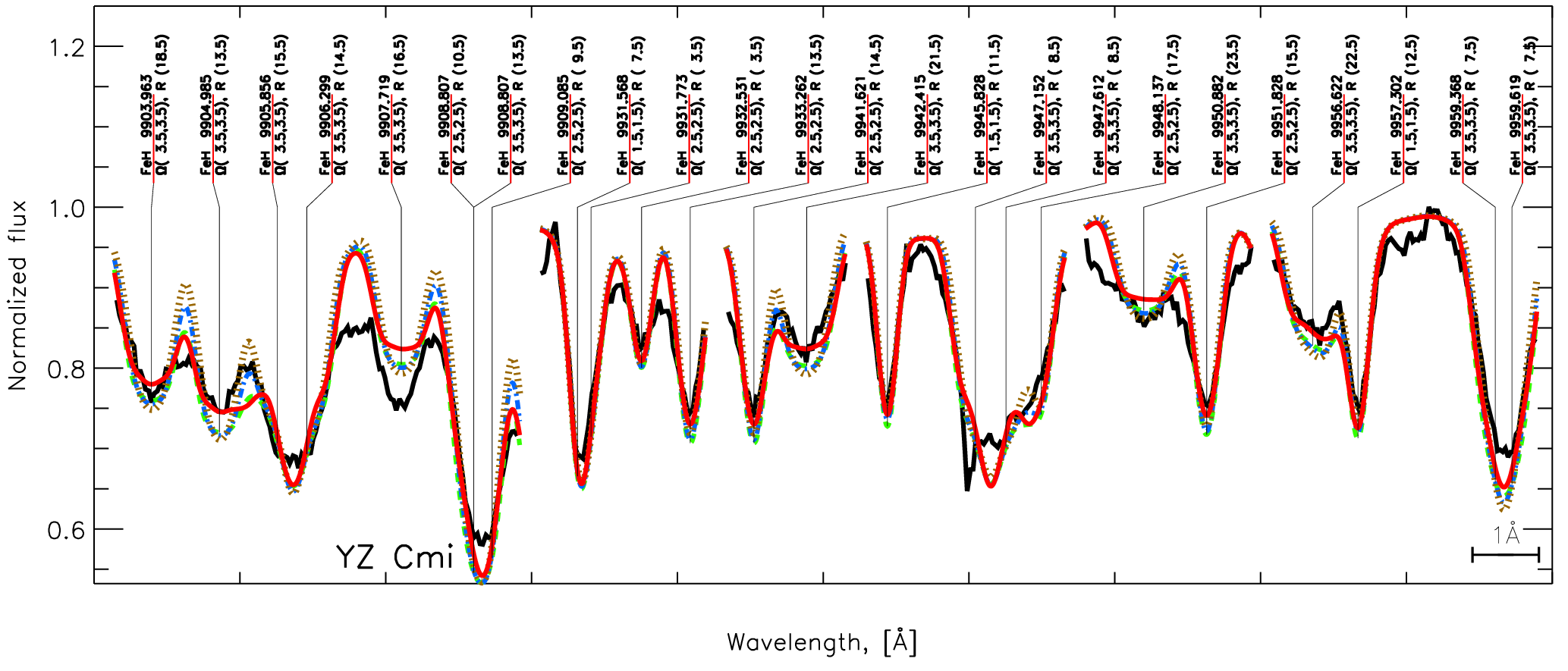}
\caption{Comparison between observed and theoretical spectra of selected M-dwarfs. 
\textbf{Upper panel:} non-active star GJ~1002, red solid line--assuming $\varepsilon_{\rm Fe}=-4.37$, blue dashed line--$\varepsilon_{\rm Fe}=-4.59$.
\textbf{Middle panel:} active star AD~Leo, green dashed~--~ $\bv=(1.7,1.7,0)$~kG, blue dash-dotted~--~$\bv=(2,0,0)$~kG, brawn dotted~--~$\bv=(2.5,0,0)$~kG,
red solid~--~ $\bv=(2.9,0,0)$~kG.
\textbf{Bottom panel:} active star YZ~CMi, green dashed~--~ $\bv=(0,4,0)$~kG, blue dash-dotted~--~$\bv=(2.5,2.5,0)$~kG, red solid~--~ $\bv=(4,0,0)$~kG,
dotted~--~$\bv=(3,0,0)$~kG.}
\label{fig:feh-dwarfs}
\end{figure*}

Having fixed g-factors for important FeH transitions, we used them to derive the magnetic field strength in selected M-dwarfs.
As an example, here we compare model predictions for active M-dwarfs AD~Leo and YZ~CMi, for which previous attempts to measure its magnetic field 
resulted in $\btimesf=2.9$~kG  and $\btimesf>3.9$~kG respectively \citep{r-and-b-2007}.
Theoretical fit to FeH lines is shown in the Fig.~\ref{fig:feh-dwarfs} together with the predictions for the non-active M$4.5$ star GJ~1002. 
In particular, magnetically sensitive lines such as FeH $9905$~\AA, $9906$~\AA, $9942$~\AA, and $9959$~\AA\ clearly 
point to the lower field modulus. That the widths of these lines are well reproduced in the sunspot spectra 
(see Fig.~\ref{fig:spot-feh}) allows us to consider them as important indicators of the mean field intensity. In particular,
increasing $\b$ results in the appearance of the characteristic feature owing to the crossed $\sigma$-components of the two
FeH lines at $9906$~\AA. Overlaid, these components give rise to the absorption feature \textit{which is not seen in the observed spectra}.
Consequently, weaker fields are needed to keep these lines separated.

\begin{table}
\caption{Atmospheric parameters of investigated M-dwarfs.}
\label{tab:stellar}
\begin{footnotesize}
\begin{center}
\begin{tabular}{lcccrrr}
\hline
Name    & Spectral      & $\teff$ & \vsini & \multicolumn{2}{c}{$\b_{\mathrm{m}}$} & \multicolumn{1}{c}{$\btimesf$}\\
        &  type         & (K)     & (\kms) & \multicolumn{2}{c}{(kG)}              & \multicolumn{1}{c}{(kG)}\\
        &               &         &        & atoms             & FeH               &\\
\hline
Sunspot & --            & 4000    & 0.0    & $2.7$             & $2.7$             & $2.7^{(1)}$\\
GJ~1002 & M$5.5$        & 3100    & 2.5    & $0$               & $0$               & --\\
GJ~1224 & M$4.5$        & 3200    & 3.0    & $\approx2$        & $1.7-2$           & $2.7^{(2)}$\\
YZ~Cmi  & M$4.5$        & 3300    & 5.0    & $3-4$             & $3-3.5$           & $>3.9^{(2)}$\\
EV~Lac  & M$3.5$        & 3400    & 1.0    & $3-4$             & $3-3.5$           & $\sim3.9^{(3)}$\\
AD~Leo  & M$3.5$        & 3400    & 3.0    & $2-3$             & $2-2.5$           & $\sim2.9^{(2)}$\\
\hline
\end{tabular}
\end{center}
$\b_{\mathrm{m}}$~--~mean surface magnetic field\\
$\btimesf$~--~results of previous investigations\\\\
(1)~--~\citet{wallace1998}\\
(2)~--~\citet{r-and-b-2007}, scaled from (3)\\
(3)~--~\citet{jk-valenti1996}
\end{footnotesize}
\end{table}

Table~\ref{tab:stellar} gathers the main results of the present study including several other M-dwarfs which were
analyzed applying the same approach. An average surface magnetic field resulting from the analysis
of atomic and FeH lines are shown separately. For GJ~1224, the measurements of the magnetic field from atoms were
based on Ti~I lines located in the region $10\,400-10\,800$~\AA\AA, while for AD~Leo, YZ~CMi, and EV~Lac results are
from the  \ion{Fe}{i}~$8468$~\AA\ line (the same way as presented in \citet{jk-valenti1996}).
A large scatter resulting from the analysis of the \ion{Fe}{i}~$8468$~\AA\ line
is due to the uncertainties of fitting blue and red wings of this line, 
which seem to require different $\b$. The low quality of the data (low resolution and signal-to-noise ratio)
can also introduce uncertainties in the fitting procedure, and thus it is difficult to draw strict conclusions
without further more accurate analysis. In addition, fitting the spectra of magnetic stars we assumed the magnetic field
to be homogeneous everywhere on the stellar surface (corresponding filling factor $f=1$), 
while previous studies assumed the stellar surface
to consist of combinations of magnetic ($f<1$) and non-magnetic ($f=0$) regions, but with the former being represented
by radial magnetic field only, i.e ignoring possible horizontal vector components of the magnetic field.
We refer the interested reader to the work of \citet{mag-cool} for more
details.

\section{Summary}

In this study we developed an approach
of modelling the Zeeman splitting in FeH lines of Wing-Ford $F^4\,\Delta-X^4\,\Delta$ band
which then applied to measure magnetic fields in selected M-dwarfs.
The main result of this study is that the magnetic field strengths derived from FeH lines are $15-30$\%\ lower than results presented in
\citet{r-and-b-2007}, which are based on atomic line analysis scaled from \citet{jk-valenti2000}.
The estimates of the magnetic field modulus from the Fe~I~$8468$~\AA\ line seem to be systematically higher 
than those from FeH lines. This needs further more extensive investigation.

\begin{acknowledgements}
We thank Dr. Bernard Leroy for his kind advice and help with the MZL routines. 

This work was supported by the following grants: Deutsche Forschungsgemeinschaft (DFG)
Research Grant RE1664/7-1 to DS and Deutsche Forschungsgemeinschaft under DFG RE 1664/4-1 and NSF
grant AST07-08074 to AS. SW acknowledges financial support from the DFG
Research Training Group GrK - 1351 "Extrasolar Planets and their host stars".
OK is a Royal Swedish Academy of Sciences Research Fellow supported 
by grants from the Knut and Alice Wallenberg Foundation and the Swedish Research Council.
We also acknowledge the use of electronic databases (VALD, SIMBAD, NASA's ADS).
This research has made use of the Molecular Zeeman Library (Leroy, 2004).
\end{acknowledgements}

\bibliography{reiners_a}

\end{document}